\documentclass[nofootinbib,aps,preprint]{revtex4}
\usepackage{graphicx,amssymb}

\newcommand{\ben}{\begin{enumerate}}
\newcommand{\een}{\end{enumerate}}
\newcommand{\bit}{\begin{itemize}}
\newcommand{\eit}{\end{itemize}}

\newcommand{\bea}{\begin{eqnarray}}
\newcommand{\eea}{\end{eqnarray}}
\newcommand{\beqa}{\begin{eqnarray}}
\newcommand{\eeqa}{\end{eqnarray}}
\newcommand{\beq}{\begin{equation}}
\newcommand{\eeq}{\end{equation}}
\newcommand{\bay}{\begin{array}}
\newcommand{\eay}{\end{array}}

\def\gsim{\ \rlap{\raise 3pt \hbox{$>$}}{\lower 3pt \hbox{$\sim$}}\ }
\def\lsim{\ \rlap{\raise 3pt \hbox{$<$}}{\lower 3pt \hbox{$\sim$}}\ }

\def\ord{{\cal O}}
\def\lt{\left}
\def\rt{\right}
\def\half{\frac{1}{2}}

\def\qp2{\lt|\frac{q}{p}\rt|^2}
\def\pq2{\lt|\frac{p}{q}\rt|^2}
\def\Af2{\lt|A_{Bf}\rt|^2}
\def\Abf2{\lt|\bar A_{Bf}\rt|^2}
\def\Afb2{\lt|A_{B\bar f}\rt|^2}
\def\Abfb2{\lt|\bar A_{B\bar f}\rt|^2}
\def\ss2{s_{12}}
\def\ss3{s_{13}}

\def\bra#1{{\langle #1|}}
\def\ket#1{{| #1\rangle}}

\def\Nt{{\tilde N}}
\def\Nd{{\tilde N^\dagger}}
\def\Lt{{\tilde L}}
\def\Lb{{\bar L}}
\def\Ld{{L^\dagger}}
\def\hb{{\bar h}}

\def\st{s}
\def\ct{c}

\def\ss{\tilde s}

\def\nav{{\langle n_f\rangle}}
\def\nbav{{\langle \bar n_f\rangle}}
\def\sa{{\sin(\alpha-\phi)}}
\def\ca{{\cos(\alpha-\phi)}}

\def\eg{{\it e.g. }}
\def\ie{{\it i.e. }}
\def\NN{{\cal N}}
\def\NNt{{\tilde\NN}}
\def\LL{{\cal L}}

\def\bracket#1#2 {\mathinner{\langle{#1}|{#2}\rangle}}

\def\slp {{\tilde L}}
\def\aslp {{\tilde L^{\dag}}}

\def \ep {\epsilon}
\def \epp {{\varepsilon}}

\def \al {{\alpha}}

\def\no{{\nonumber}}

\begin{document}

\vspace*{3cm}

\title{The (ir)Relevance of Initial Conditions in Soft Leptogenesis}

\def\addtech{Department of Physics,
Technion -- Israel Institute of Technology,\\
Haifa 32000, Israel\vspace*{1cm}}

\author{Omri Bahat-Treidel}\affiliation{\addtech}
\author{Ze'ev Surujon}\affiliation{\addtech}

\begin{abstract}
We explore how the initial conditions affect the final
lepton asymmetry in Soft Leptogenesis.
It has been usually assumed that the initial state is a statistical
mixture of sterile sneutrinos and anti sneutrinos with equal
abundances.  We calculate the lepton asymmetry due to the most general
initial mixture.
The usually assumed equal mixture produces a small, but sufficient,
lepton asymmetry which is proportional to the ratio of the
supersymmetry breaking scale over the Majorana scale.
A more generic mixture, still with equal contents of sneutrinos and anti sneutrinos,
yields an unsuppressed lepton asymmetry.  Mixtures of non equal
contents of sneutrinos and anti sneutrinos result in a large lepton
asymmetry too.
While these results establish the robustness of Soft Leptogenesis
and other mixing based mechanisms, they also expose their
lack of predictive power.
%
%

\end{abstract}
\maketitle

\section{Introduction}
The accepted paradigm in baryogenesis relies on generating
baryon asymmetry dynamically from symmetric initial conditions.
Such generation may occur if there are processes which satisfy
the Sakharov conditions~\cite{sakharov}, namely, they violate baryon
number, they violate C and CP, and they occur out of equilibrium.
In Leptogenesis scenarios~\cite{lepto}, a lepton asymmetry is
generated at a high temperature, and is partially transformed into a
baryon asymmetry by the $(B-L)$-violating sphalerons.
The $L$, C and CP violation may come from various sources, such as
the decays of sterile neutrinos~\cite{lepto,lepto2} or
sneutrinos~\cite{sacha,slepto}, the mixing between them
(Resonant Leptogenesis~\cite{resonant}
and Soft Leptogenesis~\cite{soft,soft1,soft2}), or
interference effects~\cite{newways}.
The third condition is satisfied by the decaying of the heavy particles
(sterile neutrinos and sneutrinos), which occurs out of equilibrium once the
temperature becomes low enough.

For simplicity, we limit ourselves to Soft Leptogenesis, which is
based on the minimal supersymmetric standard model plus a sterile
neutrino superfield $N$.  In this framework, the soft supersymmetry
breaking parameters are responsible for the CP violation~\footnote{
We should mention here that the idea of using supersymmetry breaking
terms for Leptogenesis was raised already in an earlier work~\cite{boub}
in the context of low scale Leptogenesis.}.
The leading contribution~\cite{soft,soft2} comes from CP violation in the mixing
between the sneutrino ($\Nt$) and the anti-sneutrino ($\Nd$), and
therefore it is present already in a model with one generation of leptons.

The lepton asymmetry produced in this mechanism depends on the state
of the \mbox{$\Nt-\Nd$} system before decaying.
It has been usually assumed in the literature
that this initial state is an equal incoherent mixture of $\ket{\Nt}$
and $\ket{\Nd}$.
Although this seems to be a reasonable assumption,  the actual initial state
could be different, since it depends
on the couplings of the sneutrinos in the high energy theory, which determine
both the details of the sneutrino production and the physics responsible for the decoherence of the plasma.
For example, the sneutrinos may be produced non-thermally by direct inflaton decay~\cite{sacha,infl-decay}.
In this case, the coupling to the inflaton is needed
in order to determine the plasma state just before Leptogenesis occurs;
In Grand Unified Theories, the initial state would depend on the way the neutrino superfield is
embedded in the matter multiplet;
In models which explain the flavor hierarchy, the coupling to the new fields (\eg flavons) and the
sneutrino representation under the flavor group could largely affect the initial state, and so on.
Previous discussions of the initial state can be found for example in~\cite{sacha,covi,rangarajan,infl-decay}.

In this work, we give a general treatment of this issue, by considering a generic initial state.
We find that the resulting lepton asymmetry is generally enhanced 
compared to the one resulting from the usually assumed initial state.
As particular cases, we consider a mixture of ``CP eigenstates''
$\ket{\Nt}\pm e^{i\phi}\ket{\Nd}$.
As in meson mixing, this labeling is appropriate in the presence of small CP violation.
These mixtures have, as in the case of the usually assumed state, equal expectation values of $\Nt$ and $\Nd$.
Such mixtures yield an enhanced asymmetry too.
A motivation for considering such mixtures would be an approximate CP conservation of the high energy theory.
Another particular case is a mixture of the Hamiltonian eigenstates which, if the CP violation is small,
are very similar to the Hamiltonian eigenstates mentioned above.
Such mixtures would be motivated by assuming that the quantum decoherence is due to the same
Hamiltonian, so that only the energy eigenstates are projected in.
This mixture also yields an enhanced lepton asymmetry.
%
%
%

Since our calculations rely on CP violation in mixing, we expect them
to be valid in any mixing based Leptogenesis, \eg Resonant
Leptogenesis~\cite{resonant}.

\section{Definitions and Approximations}  \label{defs}
As stated above, in this work we consider the Soft Leptogenesis
scenario~\cite{soft,soft1}, which employs the minimal supersymmetric
standard model plus singlet neutrino chiral superfields.
Since Soft Leptogenesis requires only one generation, the relevant part
of the superpotential is
\beq
   W = \half MNN+YLNH_u,
\eeq
where we use the notation of~\cite{soft}.
Soft supersymmetry breaking is introduced by
\beq
   {\cal L}_{\rm soft}=B\Nt\Nt+A{\tilde L}\Nt H_u,
\eeq
where $\Nt$ is the singlet sneutrino.
The soft breaking parameters are assumed, as usual, to be highly suppressed:
\beq
   \frac{B}{M^2}\ll 1, \quad \frac{A}{M}\, \ll 1.
\eeq
Yet, they may have a significant
effect proportional to the CP violating phase
\mbox{$\varphi\equiv\arg(AMB^*Y^*)$}.
This weak phase shows up in decays of $\Nt$ and $\Nd$, in the mixing
between them and in the interference between decays with and without
oscillations.  In most of the parameter space, though, the largest
effect is due to CP violation in mixing.

In analogy to meson mixing and decay, we define
\beq
   x\equiv\frac{\Delta m}{\Gamma},\quad
   y\equiv\frac{\Delta\Gamma}{2\Gamma},
\eeq
where $\Delta m$ and $\Delta\Gamma$ are the mass and width differences
of the mass eigenstates
\beq
   \ket{\Nt_{L,H}}=p\ket{\Nt}\pm q\ket{\Nd},\qquad
   \frac{q}{p}\equiv Re^{i\phi}.
\eeq
Note that $\phi$ depends on convention.
In particular, in the convention where the decay amplitudes are real, we have
$\phi=-\varphi$.
We also define
\beq
   \epsilon\equiv 1-R^2,
\eeq
which signals CP violation in mixing.  Note that
$\epsilon$ is expected to be small,
\beq
   \epsilon=\ord(\epsilon_S/x),\qquad
   \epsilon_S\equiv \frac{m_{\rm SUSY}}{M}
\eeq
(see~\cite{newways}, note slight difference in notations).

We denote the relevant amplitudes by
\beqa    \label{amps}
   A_\Lt\equiv A\lt(\Nt\to\Lt H\rt),\qquad
   A_\Lb\equiv A\lt(\Nt\to\Lb\hb\rt),\nonumber\\
   \bar A_{\Lt^\dagger}\equiv A\lt(\Nd\to\Lt^\dagger H^\dagger\rt),
   \qquad
   \bar A_L\equiv A\lt(\Nd\to Lh\rt).
\eeqa
Following~\cite{soft}, we make the following
approximations (these approximations are relaxed
in~\cite{newways} and in the appendix of this article):
\ben
\item $A_\Lt=\bar A_{\Lt^\dagger}$.
\item $A_\Lb=\bar A_L$.
\item All the two body decay amplitudes are negligible except those in
Eq.~(\ref{amps}).
\item We neglect three body decays (this is justified in~\cite{newways}).
\een
The lepton asymmetry due to the usually assumed initial state
(a mixture of $\ket{\Nt}$ and $\ket{\Nd}$ with equal abundances)
is found in~\cite{soft,soft1} to be
\beq   \label{common}
   \varepsilon=\frac{\NNt\lt|A_\Lt\rt|^2-\NN\lt|A_\Lb\rt|^2}
   {\NNt\lt|A_\Lt\rt|^2+\NN\lt|A_\Lb\rt|^2}
   \lt(\frac{x^2+y^2}{1+x^2}\rt)\epsilon,
\eeq
where $\NNt$ and $\NN$ are phase space factors (see~\cite{soft}).
In this work, we show how this result emerges from the more general
expression as a special case.

A further generalization would be to consider other types of CP
violation~\cite{newways}.
This requires considering gaugino mass insertions and relaxing the assumptions concerning amplitudes.
In the appendix we show some consequences of this generalization.


\section{The Initial State}   \label{initial}
Our Hilbert space is two dimensional, and thus an arbitrary mixture
of $k$ pure states can always be described by a $2\times 2$
density matrix.
Therefore, it is physically equivalent to some (non-equal) mixture
of two orthogonal states
\beqa \label{psis}
   \ket{\psi} &=&
   e^{i\beta}\lt(\cos\theta\ket{\Nt}
   -e^{i\alpha}\sin\theta\ket{\Nd}\rt),   \nonumber\\
   \ket{\bar\psi} &=&
   e^{i\gamma}\lt(e^{-i\alpha}\sin\theta\ket{\Nt}
   +\cos\theta\ket{\Nd}\rt),
\eeqa
which are the eigenstates of the density matrix. We further define $n$
and $\bar n$ to be the statistical abundances of $\ket{\psi}$ and $\ket{\bar \psi}$,
respectively.

The overall phase $(\beta+\gamma)$ is not observable in quantum mechanics.
Furthermore, considering statistical mixtures of $\ket{\psi}$ and $\ket{\bar\psi}$
means that we cannot measure the relative phase $(\beta-\gamma)$ as
well.  Therefore, a statistical mixture is described by three free
parameters, which are $\theta$, $\alpha$, and $n$ (the
statistical abundance of $\ket{\psi}$).  Indeed, being Hermitian and
traceless, a two dimensional density matrix has three real parameters.
In our parameterization, the density matrix takes the form
\beqa   \label{density}
   \rho = n\ket{\psi}\bra{\psi}
   +\bar n\ket{\bar\psi}\bra{\bar\psi}
   &=& \pmatrix{n\cos^2\!\theta+\bar n\sin^2\!\theta &
       -(n-\bar n)e^{-i\alpha}\cos\theta\,\sin\theta\cr
       -(n-\bar n)e^{i\alpha}\cos\theta\,\sin\theta &
       n\sin^2\!\theta+\bar n\cos^2\!\theta}
   \nonumber\\ && \nonumber\\
   &=&\half\pmatrix{1+\ct\,\Delta &
   -\st e^{-i\alpha}\Delta  \cr
   -\st e^{i\alpha}\Delta  & 1-\ct\,\Delta},
\eeqa
where
\beq
   \st\equiv\sin(2\theta),\qquad \ct\equiv\cos(2\theta),
   \qquad \Delta\equiv n-\bar n.
\eeq
We denote by $\ket{\Nt}$ and $\ket{\Nd}$ the sneutrino ``flavor'' states.
Then,
the diagonal elements are just the expectation values of the
``flavor'' state abundances ($\nav,\nbav$).
We also define
\beq    \label{flavor}
   \Delta_f \equiv \nav-\nbav = \cos(2\theta)\,\Delta
\eeq
as a measure of ``flavor'' imbalance.
A particularly interesting type of initial states is one which concerns
``flavorless'' mixtures, \ie mixtures with $\Delta_f=0$.
{}From Eqs.~(\ref{density}) and (\ref{flavor}) one can see that a
``flavorless'' mixture must be equivalent to either of the following:
\ben
\item
A mixture of (any) two orthogonal states with equal abundances
($\Delta=0$), \ie $\rho=I/2$.
A special case of such mixtures is the usual ansatz for the initial state
($\theta=\Delta=0$).
Note that once $\Delta=0$, the other mixture parameters ($\theta,\alpha$)
become physically insignificant.

\item
Mixtures of the two states: $\ket{\Nt}\pm e^{i\alpha}\ket{\Nd}$
$(\theta=45^\circ)$.
This is a two parameter family of mixtures, parametrized by
$\alpha$ and $\Delta$.
In the absence of CP violation, such a mixture would be a mixture of the CP eigenstates, analogous
to the states $\ket{K^0}\pm\ket{\overline{K^0}}$ in the neutral kaons.
Note that for a small CP violation in mixing, which is assumed throughout this work, the difference between
these states and the Hamiltonian states is very small, of order~$\epsilon\sim\epsilon_S/x$.
\een
Note that ``flavorless'' does not stand for ``with vanishing lepton number'',
as both $\Nt$ and $\Nd$ decay to both positive and negative lepton number
products.
In the next section we calculate the lepton asymmetry due
to a generic initial mixture, with an emphasis on these
``flavorless'' mixtures.


\section{Lepton Asymmetries}

If we define $\ket{\psi(t)}$ and $\ket{\bar\psi(t)}$ to be the states
which have evolved from $\ket{\psi}$ and $\ket{\bar\psi}$ at time
$t=0$, respectively, then the time integrated decay rate of
$\ket{\psi(t)}$ into a final state $\ket{f}$ is given by
\beq
   \Gamma_f=\NN\int_0^\infty\!\!\lt|\langle f|{\cal H}\ket{\psi(t)}\rt|^2 dt,
\eeq
where ${\cal H}$ is the decay Hamiltonian.
The lepton asymmetry is given by
\beq   \label{asym}
   \varepsilon \equiv\frac{
   n\lt(\Gamma_\Lt+\Gamma_L- \Gamma_{\Lt^\dagger}- \Gamma_\Lb\rt)
   +\bar n\lt(\bar \Gamma_\Lt+\bar \Gamma_L- \bar \Gamma_{\Lt^\dagger}
   -\bar \Gamma_\Lb\rt)}
   {n\lt(\Gamma_\Lt+\Gamma_L+ \Gamma_{\Lt^\dagger}+ \Gamma_\Lb\rt)
   +\bar n\lt(\bar \Gamma_\Lt+\bar \Gamma_L+ \bar \Gamma_{\Lt^\dagger}
   + \bar \Gamma_\Lb\rt)},
\eeq
where $\Gamma_X$ $\lt(\bar\Gamma_X\rt)$ are the time integrated decay rates of
$\ket{\psi(t)}$ $\lt(\ket{\bar\psi(t)}\rt)$ into a final state which
includes $X$.

We find that an initial mixture of $\psi$ and
$\bar\psi$ with abundances $n$ and $\bar n$
yields a final asymmetry
\beq  \label{generic}
   \varepsilon_\psi = \delta_{\Lt\Lb}\lt[
   a_1\Delta+\lt(b_0+b_1\Delta+b_2\Delta^2\rt)\epsilon\rt]
   + \ord\lt(\epsilon^2\rt),
\eeq
where
\beq    \label{thermal}
   \delta_{\Lt\Lb}\equiv\frac{\NNt\lt|A_\Lt\rt|^2-\NN\lt|A_\Lb\rt|^2}
   {\NNt\lt|A_\Lt\rt|^2+\NN\lt|A_\Lb\rt|^2}
\eeq
is a supersymmetry breaking factor which is of order one at a temperature of $T\sim M$
($\NNt$ and $\NN$ are phase space factors), and
\beqa   \label{generic1}
   a_1 &=& -\frac{1-y^2}{1+x^2}\,
   \lt[\frac{\ct-\sa\st x}{1+\ca\st y\Delta}\rt],\nonumber\\
   b_0 &=& \frac{x^2+y^2}{2\lt(1+x^2\rt)}
   \lt[\frac{1}{1+\ca\st y\Delta}\rt]
   ,    \nonumber\\
   b_1 &=& \frac{\ca\st y}{2\lt[1+\ca\st y\Delta\rt]},  \nonumber\\
   b_2 &=& \half \lt(\frac{1-y^2}{1+x^2}\rt)^2
   \frac{\ct-\sa\st x}{\lt[1+\ca\st y\Delta\rt]^2}
   \lt[\sa\st x+\ct\frac{x^2+y^2}{1-y^2}\rt].
\eeqa
The asymmetry due to the usually assumed initial mixture is recovered
by substituting $\Delta=\theta=0$.
Note that the coefficients $a_1,b_0,b_1$ and $b_2$ depend on $\Delta$.
An important observation is the absence of the coefficient  ``$a_0$'', which
means that the dynamical contribution ($\Delta=0$) to the asymmetry
is proportional to the parameter of CP violation in mixing $\epsilon$.
This is due to the approximations we have made in the previous section.
Since $|\Delta|< 1$, the most important coefficients are $a_1$ and $b_0$.
While $b_0$ determines the asymmetry for $\Delta=0$, $a_1$ is dominant
if $|\Delta|\gg\epsilon$.

We now turn to the special cases of flavorless initial mixtures.
First, consider the usual ansatz, namely a mixture of $\ket{\Nt}$ and
$\ket{\Nd}$ with equal abundances.
The resulting asymmetry,
\beq   \label{usual-asym}
   \varepsilon_{\rm eq}=\half\delta_{\Lt\Lb}
   \lt(\frac{x^2+y^2}{1+x^2}\rt)\epsilon,
\eeq
is given by substituting $\theta=\Delta=0$ in Eqs.~(\ref{generic}-\ref{generic1}).
This agrees with~\cite{soft,soft1}.
Recall that this ansatz corresponds to an equal mixture of $\ket{\Nt}$
and $\ket{\Nd}$ or of any other orthogonal pair of states, as explained in
the previous section.

As discussed in the previous section, a completely distinct flavorless initial
state is given by $\theta=45^\circ$.
Such a mixture is composed of the approximate CP eigenstates.
Note that this basis of states deviates from the Hamiltonian eigenbasis by $\ord(\epsilon)$.
If the high energy theory is CP conserving, such a mixture is expected to be produced.
For this special case, Eq.~(\ref{generic1}) yields
\beqa
   a_1 &=& \frac{1-y^2}{1+x^2}
   \lt[\frac{\sa x}{1+\ca y\Delta}\rt],\nonumber\\
   b_0 &=& \frac{x^2+y^2}{2\lt(1+x^2\rt)}
   \lt[\frac{1}{1+\ca y\Delta}\rt],    \nonumber\\
   b_1 &=& \frac{\ca y}{2\lt[1+\ca y\Delta\rt]},  \nonumber\\
   b_2 &=& -\half \lt(\frac{1-y^2}{1+x^2}\rt)^2
   \lt[\frac{\sa x}{1+\ca y\Delta}\rt]^2.
\eeqa
Unlike the former case, here $a_1\neq 0$, and thus the lepton asymmetry is enhanced by $\ord(1/\epsilon)$
compared to the former case.
Note that $a_1$ does vanish for $\alpha=\phi$.
Concerning an unbalanced
mixture of the flavor states, there is an interesting observation:
Substituting $\theta=0$ in Eq.~(\ref{generic1}) we obtain
\beqa   \label{flavor-mixture}
   a_1 &=& -\frac{1-y^2}{1+x^2},   \nonumber\\
   b_0 &=& \frac{x^2+y^2}{2\lt(1+x^2\rt)},    \nonumber\\
   b_1 &=& 0,  \nonumber\\
   b_2 &=& \frac{\lt(1-y^2\rt)\lt(x^2+y^2\rt)}{2\lt(1+x^2\rt)^2}.
\eeqa
Note that $b_1$ vanishes and that both $a_1$ and $b_2$ are proportional
to $1-y^2$.
Therefore, at least to order $\epsilon$, all the contribution
due to the initial asymmetry ($\Delta$ terms) is completely washed out
once one of the mass eigenstates is very long lived.
Actually, taking the $y^2\to 1$ limit recovers the asymmetry~(\ref{common})
of the usually assumed initial state.
Namely, to order $\epsilon$, if one of the states is very long lived compared with the other, the produced asymmetry does not depend
on the initial asymmetry.
This effect is not significant for Soft Leptogenesis if we assume that $y=\ord(\epsilon_S)\ll 1$ (see~\cite{newways}).
It would be significant in models where $y\approx 1-\ord(\epsilon)$.
It might be relevant in other mixing driven mechanisms of asymmetry production.

So far we have discussed only mixtures of two orthogonal states.
However, as argued in the previous section, every conceivable mixture
(\eg a mixture of two non orthogonal states, or a mixture of three
states, etc.)  is always equivalent to a mixture of two orthogonal
states such as the one given by Eq.~(\ref{psis}).  Thus, given a
generic mixture, we would like to find the equivalent orthogonal
mixture.  This can be done by diagonalizing the density matrix.  The
two orthogonal states $\ket{\psi}$ and $\ket{\bar\psi}$ are simply the
eigenvectors, and the abundances $n$ and $\bar n$ are given by the
corresponding eigenvalues.

A particularly simple example for the above is a mixture of the Hamiltonian
eigenstates $\ket{\Nt_L}$ and $\ket{\Nt_H}$, where $L,H$ stand for {\it light,heavy} respectively.
The lepton asymmetry due
to such a mixture is
\beq
   \varepsilon_M=\frac{1}{4}\delta_{\tilde L\bar L}(2\Delta+\epsilon),
\eeq
where $\Delta$ is the difference between the abundances of $\ket{\Nt_L}$
and $\ket{\Nt_H}$.
This result is most easily obtained by directly calculating
Eq.~(\ref{asym}), rather than by diagonalizing the density matrix, as
described above.  In any case, this mixture has $\Delta_f\neq 0$,
owing to the fact that when $|q/p|\neq 0$, both $\ket{\Nt_L}$ and
$\ket{\Nt_H}$ have the same asymmetric imbalance between $\ket{\Nt}$ and $\ket{\Nd}$.
This is analogous to CP violation in meson mixing, where the Hamiltonian eigenstates (\eg $K_{S,L}$)
have the same imbalance as the one between the weak eigenstates (\eg $K^0,\bar K^0$ accordingly).
Note that if the plasma evolves according to the same Hamiltonian, then indeed one would expect
decoherence effects to project the system on the Hamiltonian basis.


\section{Summary}
We have obtained the final lepton asymmetry due
to the most general initial state in the Soft
Leptogenesis scenario~\cite{soft,soft2,newways}.
We have found that a finite lepton asymmetry is generally produced in Soft Leptogenesis,
which we interpret as robustness of the Soft Leptogenesis idea against variations in the high
energy theory.
However, since the actual amount of this asymmetry depends strongly on the initial state,
it seems that the Soft Leptogenesis framework loses some of its predictive power.
Defining $\Delta_f\equiv \langle n_f\rangle-\langle \bar n_f\rangle$ to be the difference between the expectation values of
the sneutrino and anti-sneutrino initial abundances,
Eqs.~(\ref{generic}) and (\ref{generic2}) demonstrate that the most
general initial state yields both a $\,\Delta_f$-independent contribution
and an $\epsilon_S$-independent one, where
$\epsilon_S\equiv m_{\rm SUSY}/M_N$.
Since these two terms
are not related, we may say that 
a non vanishing final asymmetry is a generic feature of Soft Leptogenesis,
regardless of the type of CP violation involved.

Initial states which are ``flavorless'' ($\Delta_f=0$) must either have $\rho=I/2$
(as in the commonly assumed initial state),
or be equivalent to any mixture of the ``CP states'' ($\theta=45^\circ$).
While in the former case the asymmetry is suppressed by $\epsilon_S$,
in the latter case it is proportional only to $\Delta$, the difference between
the abundances in the mixture.

Another observation is that for an unbalanced initial state ($\Delta_f\neq 0$),
CP violation in mixing between the heavy sneutrinos
washes out part of the initial asymmetry in the sneutrino system.
If the mixture is equivalent to a mixture of the flavor states,
then Eq.~(\ref{flavor}) shows that this washout gets
strong when one of the mass eigenstates becomes very long lived
compared to the other ($\Delta\Gamma\approx 2\Gamma$).
In this limit the system behaves in a way which is similar to the usually assumed case.

As a final remark, we note that the initial state may be of relevance
also with regard to other issues, such as:
Resonant Leptogenesis~\cite{resonant},
flavor~\cite{flavor} and washout mechanisms~\cite{washout},
thermal considerations~\cite{temp}, and scattering
effects~\cite{scattering}.
Moreover, as noted in~\cite{qboltzmann}, there are effects from
quantum Boltzmann equations such as memory effects due to
interference, which are enhanced in resonance and mixing induced
Leptogenesis.
Further research should be done in order to fully understand the relation
between these effects and the initial conditions.

\appendix
\section{Other Sources of CP Violation} 

In the text above we have considered $\Nt-\Nd$ mixing as the only source of CP violation.
However, as shown in~\cite{newways}, it may not be the only effect once gaugino
masses and interactions are taken into account.
New contributions arise, which are related to CP violation in decay and in the
interference between decays in the presence of $\Nt-\Nd$ oscillations.
These contributions may even be the leading effect for some regions in the parameter space~\cite{newways}.
Considering a general initial state, we expect also these contributions to have an unsuppressed piece
proportional to $\Delta$.
We verify this behavior of the lepton asymmetry for ``flavorless'' initial mixtures
by calculating the new contributions.
We expect similar results to hold in the general case.
The calculation is carried out along the same lines as in the previous section,
namely: obtaining the new integrated decay rates and substituting them in
Eq.~(\ref{asym}).

We start with some definitions.  Following \cite{newways}, we add to the
Lagrangian a mass term for the Wino,
\beq
  \LL = m_2\tilde W^a\tilde W^a + c.c.,
\eeq
where $a=1,2,3$ is a weak isospin index.
As in~\cite{newways}, we ignore the Bino since it would have similar effects.
Note that since $m_2$ is generally
complex, there is one more weak phase in the Lagrangian,
$\phi_W\equiv \arg(m_2MB^*)$.
Other useful definitions are
\beqa
   \sin\delta_s\equiv\sin\frac{\arg\lt(\lambda_\Lt\lambda_\Ld^{-1}\rt)}{2}
   =\ord(\alpha_2),\nonumber\\
   \sin\delta_f\equiv\sin\frac{\arg\lt(\lambda_\Lb\lambda_L^{-1}\rt)}{2}
   =\ord(1).
\eeqa
We also relax the approximation $A_\Lb=\bar A_L$, and consider
non vanishing values for $\bar A_\Lt,\, A_{\Lt^\dagger}$ and $A_L,\, \bar A_\Lb$.
We also note that according to the assumptions of~\cite{newways}, it follows that
\beq
   \epsilon\sim\epsilon_S/x,
\eeq
where $\epsilon_S\equiv m_{\rm SUSY}/M$.
More details can be found in~\cite{newways}.

Next we calculate the lepton asymmetries for the flavorless ``CP'' mixtures $(\theta=45^\circ)$.
In what follows, we ignore $\ord(1)$ factors which are similar to $\delta_{\tilde L\Lb}$
defined in Eq.(\ref{thermal}).
Calculating the integrated decay rates and substituting in Eq.~(\ref{asym}),
we obtain a lepton asymmetry of the form
\beq     \label{generic2}
   \varepsilon=a_1\Delta+\lt(b_0+b_1\Delta+\ldots\rt)\epsilon_S
   +\lt(c_0+c_1\Delta+\ldots\rt)\epsilon_S^2+\ord\lt(\epsilon_S^3\rt),
\eeq
where:
\ben

\item
The term $a_1\Delta$ is the expected piece due to generalizing the initial state.
We find that
\beq
   a_1 = -\frac{x\sin(\alpha-\phi)}{2(1+x^2)}.
\eeq

\item
The term $b_0\epsilon_S$ is given by Eq.(\ref{usual-asym}), confirming the result of~\cite{soft}.
This contribution is due to CP violation in mixing.

\item
The term $c_0\epsilon_S^2$ can be divided into various contributions according to
the type of CP violation involved:
\bit
\item Another contribution from CP violation in mixing:
\beqa    \label{mix}
    \epp^{m}_{2} &=&  \frac{\epsilon x}{2(1+x^2)}
                    \Bigg[
                    \left(
                        \left|\frac{\bar A_{\slp}}{A_{\slp}} \right|
                        +
                        \left|\frac{A_{\aslp}}{\bar A_{\aslp}} \right|
                    \right)
                    \sin{\delta_s}\cos{\phi} -
                \left(
                        \left|\frac{ A_{L}}{\bar A_{L}} \right|
                        +
                        \left|\frac{\bar A_{\bar L}}{A_{\bar L}} \right|
                    \right)
                    \sin{\delta_f}\cos{\phi_W}
                    \Bigg] \no \\
            &=& \ord\lt(\frac{\alpha_2 x\ep_S^2}{1+x^2}\rt)
\eeqa

\item CP violation in decay:
\beqa      \label{decay1}
    \epp^{d} &=&
    \frac{y}{2}
    \left( \left| \frac{\bar A_{\slp}}{A_{\slp}} \right|
        - \left| \frac{ A_{\aslp}}{\bar A_{\aslp}} \right|
    \right)
    \cos{\delta_s} \cos{\phi} = \ord\lt(\al_2 \ep_S^2\rt).
\eeqa
Note that this contribution vanishes in the absence of mixing, since it
is proportional to $y$.
The fact that there is no contribution due to pure CP violation in decay is because
such a pure direct CP violation term,
\beq    \label{decay2}
    \epp_{2}^d = \frac{{\cal N}_s \left( |A_{\aslp}|^2 - |\bar A_{\slp}|^2 \right)
                    +
                    {\cal N}_f \left( |\bar A_{L}|^2 - |A_{\bar L}|^2 \right)}
                    {{\cal N}_s \left(|A_{\slp}|^2 + |\bar A_{\aslp}|^2 \right)
                            +{\cal N}_f \left(|\bar A_{L}|^2 + |A_{\bar L}|^2
                            \right)},
\eeq
vanishes to $\ord(\epsilon_S^2)$ due to the equality
\beq  \label{equality}
   |A_{\slp}|=|\bar A_{\aslp}|,\qquad
   |A_L|=|\bar A_{\bar L}|
\eeq
which holds to $\ord(\epsilon_S^2)$, and due to the CPT theorem, which asserts that
\beq   \label{CPT}
   \NN_s(|A_{\slp}|^2+|A_{\aslp}|^2)+\NN_f(|A_L|^2+|A_{\Lb}|^2)
   =\NN_s(|\bar A_{\slp}|^2+|\bar A_{\aslp}|^2)+\NN_f(|\bar A_L|^2+|\bar A_{\Lb}|^2).
\eeq

\item CP violation in the interference between two decays differing by mixing:
\beqa       \label{inter}
\epp^i &=&
    \frac{y}{2}
        \Bigg[
        \left( \left| \frac{\bar A_{\slp}}{A_{\slp}} \right|
        + \left| \frac{ A_{\aslp}}{\bar A_{\aslp}} \right|
    \right)
    \sin{\delta_s} \sin{\phi}
    -
    \left( \left| \frac{\bar A_{\bar L}}{A_{\bar L}} \right|
        + \left| \frac{ A_{L}}{\bar A_{L}} \right|
    \right)
    \sin{\delta_f} \sin{\phi_W}
    \Bigg] \no \\
    & =& \ord\lt(\al_2 \ep_S^2\rt).
\eeqa%

\item CP violation from combined sources:
\beq       \label{comb}
     \epp^{mdi} =  -\frac{\epsilon x}{2(1+x^2)}
                    \left(
                        \left|\frac{\bar A_{\slp}}{A_{\slp}} \right|
                        -
                        \left|\frac{A_{\aslp}}{\bar A_{\aslp}} \right|
                    \right)
                    \cos{\delta_s} \sin{\phi}
                    = \ord\left(\frac{\al_2 x\ep_S^2}{1+x^2}\right).
\eeq%

\eit
This confirms the results of~\cite{newways}, up to a minus sign in
Eqs.(\ref{decay1},\ref{inter}).

\item
The coefficients $b_1,c_1,\ldots$ signify contributions proportional to
both $\Delta$ and $\epsilon_S$.
Since $\epsilon_S\ll 1$, their effect is negligible compared to that of $a_1$.
\een
This demonstrates that Soft Leptogenesis generally predicts a lepton asymmetry for
any initial state, irrespective of the type of CP violation involved.


\acknowledgments
We thank Mu-Chun Chen for asking the question which led to
the investigation of initial state effects.
We also thank Yuval Grossman and Yossi Nir for useful discussions and for
reading the manuscript.

\end{document}